\newif\ifstoc
\stocfalse    % Uncomment for arXiv / personal manuscript release

\ifstoc
  \documentclass[anonymous,review]{acmart}
\else
  \documentclass[manuscript]{acmart} 
\fi

\usepackage{libertine}   % optional: text font family
\usepackage{amsmath,amssymb,amsthm}

\author{Marc Harary}
\affiliation{%
    \institution{Columbia University}
    \department{Department of Computer Science}
    \city{New York}
    \state{NY}
    \country{USA}
}
\email{mh4654@columbia.edu}
\authornote{Preferred email: \href{mailto:hararymarc2@gmail.com}{hararymarc2@gmail.com}.}

\usepackage{amsmath,amssymb,amsthm}
\usepackage{algorithm,algpseudocode}
\usepackage{booktabs,longtable}
\usepackage{blkarray}
\usepackage{mathtools}
\usepackage[most]{tcolorbox}
\usepackage{enumerate}
\usepackage{subcaption}
\usepackage[nameinlink,capitalise]{cleveref}

\newtheorem{lemma}{Lemma}
\newtheorem{theorem}{Theorem}
\newtheorem{corollary}{Corollary}
\newtheorem{proposition}{Proposition}
\newtheorem{definition}{Definition}
\newtheorem*{remark}{Remark}

\usepackage[most]{tcolorbox}
\newtcolorbox{graybox}{
  colback=gray!4,
  colframe=gray!40,
  boxrule=0.4pt,
  sharp corners,
  left=4pt,
  right=4pt,
  top=4pt,
  bottom=4pt,
  breakable,
  enhanced,
  before skip=6pt plus 2pt minus 1pt,
  after skip=6pt plus 2pt minus 1pt
}
\begin{document}

\title{A Courcelle-Type Metatheorem for Rank-Bounded Unconstrained Binary Optimization}

\begin{abstract}
We present the first uniform XP exact algorithm for unconstrained binary optimization of quadratic, polynomial, fractional, and other objectives under a single parameter, the differentially affine (DA) rank $r$. An objective $f: \{0,1\}^n \to \mathbb{R}$ has DA rank $r$ if there is a feature map $\psi: \{0,1\}^n \to \mathbb{R}^r$ such that each coordinate flip has finite gain $\Delta_{\pm e_i}f(x)=\langle v_{\pm e_i},\psi(x)\rangle+\beta_{\pm e_i}$. Our algorithm enumerates the $O((2n)^r)$ chambers of the induced hyperplane arrangement and applies a two-sided local-optimality test: a solution exists on a chamber and is unique iff $\operatorname{sign}\Delta_{+e_i}=-\operatorname{sign}\Delta_{-e_i}$ for all $i$, in which case $x_i^\star=1$ iff $\Delta_{+e_i}>0$. This yields $n^{O(r)}$ time with $O(n)$ decoding per chamber. The framework uniformly covers a wide range of nonlinear functions, including all rank-$r$ quadratics, low-Waring-rank pseudo-Boolean polynomials, finite products/ratios on positive domains, finite-basis separable sums via explicit lifts, Taylor-series approximations of analytic functions, and compositions of all the foregoing. Applications include Ising spin models, optimal experimental design, portfolio optimization, and robust statistics. Prior to our work, only specialized subcases involving sparsity, convexity, submodularity, etc. were known to be tractable. Analogous in spirit to Courcelle’s theorem (MSO on bounded treewidth graphs) and Grohe’s meta-theorems for constraint satisfaction, our result replaces logical width with analytic rank for nonlinear pseudo-Boolean optimization. % We also suggest the possibility of parallelizable, pseudo-random, and approximate implementations of our method for practical use.
\end{abstract}

\maketitle

\section{Introduction}
\subsection{Problem setting}

Many important NP-hard optimization problems consist of a nonlinear objective over the Hamming hypercube $\{0, 1\}^n$. Examples include quadratic unconstrained binary optimization (QUBO) \cite{punnen2022quadratic}, polynomial unconstrained binary optimization (PUBO), fractional unconstrained binary optimization (FUBO), nonlinear assignment and matching problems, portfolio optimization, and optimal experimental design. 

In many such problems, the gain from flipping a coordinate---the change $f(x \pm e_i) - f(x)$---depends only on a low-dimensional linear function of the current state $x$. This occurs in low-rank quadratic and polynomial forms, in fractional objectives after cross-multiplication, and other objectives.

We show that the fundamental structural parameter is not the polynomial degree of $f$, but its \emph{differential rank}: the dimension of the feature space governing all such directional gains. We identify a single predicate condition under which local optimality corresponds to membership in a low-dimensional hyperplane arrangement, making the discrete search space enumerable in $n^{O(r)}$ time and yielding the first uniform XP tractability result for nonlinear pseudo-Boolean optimization.

\subsection{Our contributions}
\begin{itemize}
  \item \textbf{Differential affineness (DA rank).}
  We parameterize unconstrained pseudo-Boolean objectives by DA rank $r$: there exists $\psi\colon\{0,1\}^n\to\mathbb{R}^r$ with flip gains $\Delta_{\pm e_i}f(x)=\langle v_{\pm e_i},\psi(x)\rangle+\beta_{\pm e_i}$. This single parameter covers rank-$r$ quadratics, low-Waring polynomials, products/ratios on positive domains, affine/polynomial compositions, finite-basis separable sums, and bilinear/multilinear forms, with explicit lifts for each class.
  
  \item \textbf{Two-sided local optimality criterion.}
  We prove a chamberwise optimality test: on any chamber, a locally optimal solution exists iff $\operatorname{sign}\Delta_{+e_i}=-\operatorname{sign}\Delta_{-e_i}$ for all $i$, with unique decode $x_i^\star=\mathbf{1}\{\Delta_{+e_i}>0\}$. This correctly excludes chambers where any coordinate has gains of the same sign in both directions, resolving the boundary-behavior subtlety.
  
  \item \textbf{XP enumeration algorithm.}
  With $m=2n$ hyperplanes, the arrangement in $\mathbb{R}^r$ has $O((2n)^r)$ chambers. Enumerating them and applying the $O(n)$ optimality test yields an $n^{O(r)}$ exact algorithm with no external oracle calls.
  
  \item \textbf{Explicit constructions.}
  We provide closed-form feature maps $\psi$, covectors $\{v_{\pm e_i}\}$, and offsets $\{\beta_{\pm e_i}\}$ with rank bounds for all major function classes: polynomials (via Waring decomposition), products and ratios, compositions, separable sums, and CP-rank tensor models.
  
  \item \textbf{Applications.}
  The framework yields $n^{O(r)}$ exact algorithms for rank-$r$ QUBO, PUBO, and FUBO, as well as DA-rank formulations of problems in trimmed/robust statistics, optimal experimental design, and portfolio optimization.
  
  \item \textbf{Conceptual significance.}
  DA rank plays a Courcelle-style width role for nonlinear pseudo-Boolean optimization, reducing global optimization to sign-class enumeration in $\mathbb{R}^r$. This yields the first unified parameterized complexity theory for this domain, analogous to treewidth-based results for constraint satisfaction.
\end{itemize}

\subsection{Roadmap}

\cref{sec:prelim} fixes notation and recalls affine and central hyperplane arrangements. \cref{sec:da-xp-unconstrained} defines differential affineness and proves the main metatheorem, yielding the $n^{O(r)}$ exact algorithm. \cref{sec:objectives} provides explicit lifts and rank bounds for quadratics, low-Waring-rank polynomials, products/ratios on positive domains, affine/polynomial compositions, finite-basis separable sums, and bilinear/multilinear forms. \cref{sec:applications} instantiates the framework for trimmed/robust statistics, optimal experimental design (D/A/E-optimality), and portfolio optimization. \cref{sec:existing-paradigms} situates DA rank alongside existing parametric programming techniques and Courcelle/Grohe meta-theorems. \cref{sec:conclusion} summarizes the consequences of bounded DA rank and outlines extensions beyond the unconstrained setting.

\section{Preliminaries}
\label{sec:prelim}

\subsection{Notation}

We write $[n] := \{1,\dots,n\}$ for the index set. For $x \in \{0,1\}^n$, define its \emph{support} $\mathrm{supp}(x) := \{i : x_i = 1\}$. A \emph{pseudo-Boolean function} is any $f:\{0,1\}^n\to\mathbb{R}$. For a feasible perturbation $d$, the \emph{gain} is
\[
\Delta_d f(x) := f(x+d)-f(x).
\]
A \emph{feature map} is any function $\psi:\mathbb{R}^n\to\mathbb{R}^r$.

\subsection{Hyperplane arrangements}
\label{sec:hyperplane-arrangements}

A \emph{central hyperplane arrangement} in $\mathbb{R}^r$ is a finite set $\mathcal{A}=\{H_i\}_{i=1}^m$ with
\[
H_i := \left\{\xi \in \mathbb{R}^r : w_i^\top \xi = 0\right\},
\]
where $w_i \in \mathbb{R}^r\setminus\{0\}$. A central arrangement defines an \emph{oriented matroid} $\mathcal{M}(\mathcal{A})$ on ground set $[m]$. Each point $\xi\in\mathbb{R}^r$ induces a sign vector $\sigma(\xi)= \left(\mathrm{sign}\left(w_1^\top\xi\right),\ldots,\mathrm{sign}\left(w_m^\top\xi\right)\right)\in\{-,0,+\}^m$, and the set of all such sign vectors forms the covector set of $\mathcal{M}(\mathcal{A})$. The poset of covectors ordered by sign-refinement is isomorphic to the \emph{face lattice} $\mathcal{F}(\mathcal{A})$. The following classical result controls the combinatorial complexity of arrangements and zonotopes:
\begin{theorem}[Zaslavsky \cite{Zaslavsky1975}]
\label{thm:zaslavsky}
Let $\mathcal{A}$ be a central arrangement of $m$ hyperplanes in $\mathbb{R}^{r}$. Then
\[
|\mathcal{F}(\mathcal{A})| \le \sum_{j=0}^r \binom{m}{j} = O\left(m^r\right).
\]
In particular, the number of maximal top-dimensional faces (known as \emph{chambers} or \emph{topes}) is given by $|\mathcal{T}(\mathcal{A})| = O\left(m^{r}\right)$.
\end{theorem}

\section{Differentially affine objectives and chamber-wise readout (unconstrained)}
\label{sec:da-xp-unconstrained}

We work on the unconstrained domain $P=\{0,1\}^n$ with elementary feasible directions (EFDs)
\[
\mathcal{D}_1:=\{\pm e_i:\ i\in[n]\}.
\]
For $f:P\to\mathbb{R}$ and feasible $(x,d)$ with $d\in\mathcal{D}_1$, write the reduced cost
\[
\Delta_d f(x):=f(x+d)-f(x).
\]

\begin{definition}[Differentially affine (DA) of rank $r$]
\label{def:DA-unconstrained}
We say $f$ is \emph{differentially affine of rank $r$} with respect to $\mathcal{D}_1$ (denoted $f\in\mathcal{DA}^r(\mathcal{D}_1)$) if there exist a feature map $\psi:P\to\mathbb{R}^r$, covectors $v_d\in(\mathbb{R}^r)^*$, and scalars $\beta_d\in\mathbb{R}$ such that
\begin{equation}
\label{eq:DA-unconstrained}
\Delta_d f(x)=\langle v_d,\psi(x)\rangle+\beta_d,
\qquad \forall x\in P,\ \forall d\in\mathcal{D}_1\text{ feasible at }x.
\end{equation}
The minimal such $r$ is the \emph{differential rank} $\mathrm{rank}_{\mathrm{DA}}(f;\mathcal{D}_1)$.
\end{definition}

Let $\mathcal{A}(\mathcal{D}_1)$ be the affine hyperplane arrangement in $\mathbb{R}^r$ with one hyperplane $H_d:=\{\xi\in\mathbb{R}^r:\ \langle v_d,\xi\rangle+\beta_d=0\}$ per direction $d\in\mathcal{D}_1$. Write $\mathcal{T}(\mathcal{A})$ for its full-dimensional cells (chambers). For $\xi\notin\bigcup_d H_d$, define the \emph{EFD sign vector}
\[
\sigma(\xi):=\left(\operatorname{sign}(\langle v_d,\xi\rangle+\beta_d)\right)_{d\in\mathcal{D}_1}\in\{-,+\}^{2n},
\]
which is constant on each chamber $C$; write $\sigma(C)$.

\begin{lemma}[Chamber-constancy of EFD signs]
\label{lem:chamber-constancy-unconstrained}
If $f\in\mathcal{DA}^r(\mathcal{D}_1)$ and $\psi(x)\in C\in\mathcal{T}(\mathcal{A})$, then for every feasible $d\in\mathcal{D}_1$ the sign $\operatorname{sign}\Delta_df(x)$ depends only on $C$ and $d$ and equals the $d$-th entry of $\sigma(C)$.
\end{lemma}

\begin{proof}
Immediate from \eqref{eq:DA-unconstrained}.
\end{proof}

\subsection{Two-sided toggles, local stability, and exact readout}

For $i\in[n]$ and chamber $C$, introduce the two toggle signs (well-defined by \Cref{lem:chamber-constancy-unconstrained})
\[
s_i^{+}(C):=\operatorname{sign}\left(\Delta_{+e_i} f(x)\right),\qquad
s_i^{-}(C):=\operatorname{sign}\left(\Delta_{-e_i} f(x)\right)\qquad(\psi(x)\in C).
\]

\begin{definition}[Toggle-stability in a chamber]
\label{def:toggle-stability}
A vector $x^\star\in\{0,1\}^n$ with $\psi(x^\star)\in C$ is \emph{toggle-stable in $C$} if, for every $i$,
\[
x^\star_i=0\ \Rightarrow\ \Delta_{+e_i} f(x^\star)\le 0,\qquad
x^\star_i=1\ \Rightarrow\ \Delta_{-e_i} f(x^\star)\le 0.
\]
\end{definition}

\begin{lemma}[Per-coordinate feasibility, forcing, and ambiguity]
\label{lem:pc-conditions}
Fix a chamber $C$ and assume nondegeneracy $s_i^\pm(C)\in\{\pm1\}$.
\begin{enumerate}
\item \emph{Feasibility filter:} A toggle-stable point with $\psi(x)\in C$ exists if and only if no coordinate has $(s_i^{+},s_i^{-})=(+,+)$.
\item \emph{Forced choices:} If $(s_i^{+},s_i^{-})=(+,-)$ then any toggle-stable $x$ must have $x_i=1$; if $(s_i^{+},s_i^{-})=(-,+)$ then any toggle-stable $x$ must have $x_i=0$.
\item \emph{Ambiguity:} If $(s_i^{+},s_i^{-})=(-,-)$, then both $x_i=0$ and $x_i=1$ satisfy \Cref{def:toggle-stability} at coordinate $i$.
\end{enumerate}
\end{lemma}

\begin{proof}
(1) If $(+,+)$ occurs at $i$, then $\Delta_{\pm e_i}f(x)>0$ for any $x$ with $\psi(x)\in C$, so no $x$ is stable. Conversely, if each coordinate avoids $(+,+)$, assign $x_i$ by (2) and choose arbitrarily when $(-,-)$; the resulting $x$ meets \Cref{def:toggle-stability}.  
(2) If $(+,-)$ and $x_i=0$, then $\Delta_{+e_i}f(x)>0$ contradicts stability; hence $x_i=1$ is forced. The case $(-,+)$ is symmetric.  
(3) If $(-,-)$, then both toggles decrease $f$, so either value of $x_i$ is locally admissible.
\end{proof}

\begin{proposition}[Two-sided readout; uniqueness iff opposite signs]
\label{prop:unconstrained-readout-iff}
Fix a chamber $C$ with nondegenerate signs. Then:
\begin{enumerate}
\item \emph{Existence:} A toggle-stable point with $\psi(x)\in C$ exists iff the feasibility filter of \Cref{lem:pc-conditions}(1) holds.
\item \emph{Uniqueness:} A toggle-stable point is unique iff $s_i^{+}(C)=-s_i^{-}(C)$ for every $i$. In that case,
\[
x^\star_i(C)=\mathbf 1 \{s_i^{+}(C)=+\}\qquad(i\in[n]).
\]
\item \emph{Optimality within $C$:} Any global maximizer $x^{\mathrm{opt}}$ with $\psi(x^{\mathrm{opt}})\in C$ is toggle-stable, since an improving toggle would contradict global optimality. If uniqueness holds, then $x^{\mathrm{opt}}=x^\star(C)$.
\end{enumerate}
\end{proposition}

\begin{proof}
(1) Follows immediately from \Cref{lem:pc-conditions}(1): if any coordinate has $(s_i^+,s_i^-)=(+,+)$ then no point is toggle-stable; if none do, assigning coordinates per \Cref{lem:pc-conditions}(2) and choosing arbitrarily when $(-,-)$ yields a toggle-stable point.

(2) If some $(s_i^+,s_i^-) = (-,-)$, then both $\Delta_{+e_i}f(x)<0$ and $\Delta_{-e_i}f(x)<0$ on $C$, so both $x_i=0$ and $x_i=1$ satisfy \Cref{def:toggle-stability} at coordinate $i$, giving at least two distinct toggle-stable points. Conversely, if $s_i^+(C)=-s_i^-(C)$ for all $i$, then \Cref{lem:pc-conditions}(2) forces $x_i=1$ when $(s_i^+,s_i^-)= (+,-)$ and $x_i=0$ when $(-,+)$, yielding the unique decode $x^\star_i(C)=\mathbf{1}\{s_i^{+}(C)=+\}.$

(3) Let $x^{\mathrm{opt}}$ be a global maximizer with $\psi(x^{\mathrm{opt}})\in C$. If some coordinate admits an improving toggle, then $f$ increases, contradicting global optimality. Hence $x^{\mathrm{opt}}$ is toggle-stable. When uniqueness from (2) holds, necessarily $x^{\mathrm{opt}}=x^\star(C)$.
\end{proof}

\begin{remark}[Degeneracy and generic perturbation]
\label{rem:degeneracy}
If some $s_i^{\pm}(C)=0$, apply an infinitesimal perturbation $\beta_{\pm e_i}\leftarrow \beta_{\pm e_i}+\varepsilon\rho_{\pm e_i}$ with fixed $\rho$ and $\varepsilon\downarrow 0$ to obtain strict signs without changing the DA rank; the limit selects a maximizer under a lexicographic refinement of $f$.
\end{remark}

\subsection{Enumeration complexity and algorithmic guarantee}

Here $|\mathcal{D}_1|=2n$. By \Cref{thm:zaslavsky}, the number of chambers satisfies
\[
|\mathcal{T}(\mathcal{A})|=O\left((2n)^r\right)=n^{O(r)}.
\]

\begin{theorem}[XP meta-theorem in differential rank (unconstrained)]
\label{thm:xp-unconstrained}
Let $f\in\mathcal{DA}^r(\mathcal{D}_1)$. There is an algorithm that enumerates all $O\left((2n)^r\right)$ chambers $C\in\mathcal{T}(\mathcal{A})$ and, for each $C$:
\begin{enumerate}
\item reads the two-sided signs $\{s_i^{\pm}(C)\}_{i=1}^n$ in $O(n)$ time;
\item discards $C$ if any $(s_i^{+},s_i^{-})=(+,+)$;
\item otherwise outputs the unique toggle-stable $x^\star(C)$ when all pairs are opposite, or applies the perturbation scheme of \Cref{rem:degeneracy} to break ties.
\end{enumerate}
Among all chambers that pass the filter, return the best $x^\star(C)$. The total running time is $O\left(n\cdot(2n)^r\right)=n^{O(r)}$.
\end{theorem}

\section{Objective instantiations}
\label{sec:objectives}

\subsection{Quadratic objectives are differentially affine}
\label{subsec:quad-DA}

Let $Q\in\mathbb{R}^{n\times n}$ be symmetric with $\operatorname{rank}(Q)=r$. Fix a rank factorization
\[
Q = B^\top B,\qquad B\in\mathbb{R}^{r\times n}\ \text{full row rank}.
\]
Consider $f(x)=x^\top Q x$ on any feasible set $P\subseteq\{0,1\}^n$ and any finite EFD set $\mathcal D\subset\mathbb{Z}^n$ (e.g., $\mathcal D_{1}=\{\pm e_i\}$ for unconstrained toggles and $\mathcal D_{2}=\{\pm(e_i-e_j)\}$ for Top-$K$ exchanges).

\begin{proposition}[Quadratics are DA of rank $\boldsymbol{r}$]
\label{prop:quad-is-DA}
Define the feature map $\psi:P\to\mathbb{R}^r$ by $\psi(x):=Bx$. For each $d\in\mathcal D$, set
\[
v_d := 2B d \in \mathbb{R}^r, \qquad \beta_d := d^\top Q d \in \mathbb{R}.
\]
Then, for every feasible $(x,d)$,
\[
\Delta_d f(x)= f(x + d)-f(x) = 2d^\top Q x + d^\top Q d = \langle v_d,\ \psi(x)\rangle + \beta_d.
\]
Hence $f\in\mathcal{DA}^{r}(\mathcal D)$ with differential rank equal to $\operatorname{rank}(Q)$.
\end{proposition}

\begin{proof}
Using $Q=B^\top B$,
\[
\Delta_d f(x)=(x + d)^\top Q(x{+}d)-x^\top Qx
=2d^\top Qx + d^\top Qd
=2(Bd)^\top(Bx)+d^\top Q d
=\langle 2Bd, Bx\rangle + \beta_d.
\]
This matches the DA form with $\psi(x)=Bx$, $v_d=2Bd$, and $\beta_d=d^\top Qd$.
\end{proof}

\begin{corollary}[Affine–quadratics remain DA of rank $\boldsymbol{r}$]
\label{cor:affine-quad-DA}
For $f(x)=x^\top Q x + c^\top x + \mathrm{const}$, the same $\psi$ and $v_d$ as in \Cref{prop:quad-is-DA} yield
\[
\Delta_d f(x) = \langle v_d, \psi(x)\rangle + \underbrace{\left(d^\top Q d + c^\top d\right)}_{\beta_d},
\]
so $f\in\mathcal{DA}^{r}(\mathcal D)$. The linear term contributes only to the intercepts $\{\beta_d\}$ and does not increase differential rank.
\end{corollary}

This subsumes Ising \cite{ising1925beitrag} and Potts \cite{potts1952some} models with low-rank coupling; pairwise Markov random field \cite{sherrington1975solvable} and conditional random field \cite{lafferty2001conditional} energies; and Hopfield networks \cite{lafferty2001conditional} with low-rank $Q$.

\subsection{Polynomial objectives}
\label{subsec:poly}

Let $f:\{0,1\}^n\to\mathbb{R}$ be a pseudo-Boolean polynomial of degree at most $D$,
\[
f(x)=\sum_{t=0}^{D}\left\langle \mathcal{T}^{(t)},x^{\otimes t}\right\rangle,
\]
where each $\mathcal{T}^{(t)}$ is a symmetric order-$t$ tensor. Assume a symmetric CP/Waring factorization \cite{kolda2009tensor}
\[
\mathcal{T}^{(t)}=\sum_{\ell=1}^{k_t}\alpha_{t\ell}\left(u^{(t,\ell)}\right)^{\otimes t},
\qquad \alpha_{t\ell}\in\mathbb{R},\ u^{(t,\ell)}\in\mathbb{R}^n.
\]
Define the base linear forms $s_{t\ell}(x):=\left\langle u^{(t,\ell)},x\right\rangle$ and let
\[
r :=\sum_{t=1}^{D} tk_t,\qquad
\psi(x) := \left(s_{t\ell}(x)^{q} \right)_{\substack{t=1,\dots,D\\ \ell=1,\dots,k_t\\ q=0,\dots,t-1}} \in \mathbb{R}^{r}.
\]

\begin{proposition}[Low-rank polynomials are DA]
\label{prop:poly-dl-efd}
For any finite EFD set $\mathcal D$ and any feasible $(x,d)$ (so $x,x{+}d\in P$), there exists $w_d\in \left(\mathbb{R}^{r}\right)^*$, independent of $x$, such that
\[
\Delta_{d} f(x) = \left\langle w_d, \psi(x)\right\rangle.
\]
Hence $f\in\mathcal{DA}^{r}$ with $r=\sum_{t=1}^{D} tk_t$.
\end{proposition}

\begin{proof}
Write $f=\sum_{t,\ell} \alpha_{t\ell} s_{t\ell}(x)^{t}$. For fixed $(t,\ell)$,
\[
s_{t\ell}(x{+}d)^{t}-s_{t\ell}(x)^{t} = \sum_{j=1}^{t}\binom{t}{j} s_{t\ell}(x)^{t-j} \left\langle u^{(t,\ell)},d\right\rangle^{j}.
\]
Summing over $(t,\ell)$ with coefficients $\alpha_{t\ell}$ yields a linear combination of the coordinates $s_{t\ell}(x)^{q}$, $q=0,\dots,t-1$, with $d$-dependent coefficients; collect them into $w_d$.
\end{proof}

This subsumes any degree-$D$ pseudo-Boolean with bounded Waring ranks $\{k_t\}$ (e.g., sparse polynomial expansions \cite{blatman2011adaptive}, low-rank higher-order MRFs \cite{sherrington1975solvable}).

\subsection{Closure under products and ratios}
\label{subsec:products-ratios}

Let $\{f_s\}_{s=1}^S$ with $f_s\in\mathcal{DA}^{r_s}$ admit feature maps $\psi_s$ so that for every EFD $d$,
\[
\Delta_d f_s(x)=\left\langle w_{s,d},\psi_s(x)\right\rangle+\beta_{s,d}.
\]

\begin{proposition}[Finite products preserve DA]
\label{prop:da-products}
Define the lifted feature map
\[
\Psi_{\Pi}(x) := \bigotimes_{s=1}^S \left[\psi_s(x);\ f_s(x);\ 1 \right] \ \in\ \mathbb{R}^{R_{\Pi}},
\qquad R_{\Pi} \le \prod_{s=1}^S (r_s+2).
\]
Then for every EFD $d$ there exists $W^{(\Pi)}_d$ with
\[
\Delta_d\left(\prod_{s=1}^{S} f_s\right)(x)=\left\langle W^{(\Pi)}_d,\Psi_{\Pi}(x)\right\rangle,
\]
hence $\prod_{s} f_s \in \mathcal{DA}^{R_{\Pi}}$.
\end{proposition}

\begin{proof}
Let $A_s=f_s(x)$ and $B_s=\Delta_d f_s(x)=\langle w_{s,d},\psi_s(x)\rangle+\beta_{s,d}$. The discrete product rule
$
\prod_s(A_s{+}B_s)-\prod_s A_s=\sum_{\emptyset\neq S'\subseteq[S]} \left(\prod_{s\in S'}B_s\right)\left(\prod_{t\notin S'}A_t\right)
$
expresses the gain as a linear form in $\bigotimes_s[\psi_s;f_s;1]$.
\end{proof}

\begin{proposition}[Ratios: affine comparator; DA only in special cases]
\label{prop:ratios-affine-comparator}
Let $F(x):=P(x)/Q(x)$ with $Q(x)>0$ on $P$, where $P,Q$ are built from $\{f_s\}$ by finite sums/products. Let $\Psi_{P},\Psi_{Q}$ be their product-lifts and set $\Psi_{\mathrm{rat}}(x):=\Psi_{P}(x)\otimes\Psi_{Q}(x)$. Then for every EFD $d$ there exists $W^{(\mathrm{rat})}_d$ such that
\[
\Delta_d F(x) = \frac{P(x{+}d)Q(x)-P(x)Q(x{+}d)}{Q(x)Q(x{+}d)}
= \frac{\left\langle W^{(\mathrm{rat})}_d,\ \Psi_{\mathrm{rat}}(x)\right\rangle}{Q(x)Q(x{+}d)}.
\]
Consequently,
\[
\operatorname{sign}\left(\Delta_d F(x)\right)=\operatorname{sign}\left(\left\langle W^{(\mathrm{rat})}_d,\ \Psi_{\mathrm{rat}}(x)\right\rangle\right).
\]
In general $F\notin\mathcal{DA}^{r}$ (the denominator depends on $x$); $F\in\mathcal{DA}^{r}$ only under exceptional conditions, e.g., when for each $d$ the product $Q(x)Q(x{+}d)$ is independent of $x$ (in particular, if $Q$ is constant on $P$).
\end{proposition}

\begin{proof}
Expand the numerator via \Cref{prop:da-products} at $x$ and $x{+}d$ to obtain a linear form in $\Psi_{\mathrm{rat}}(x)$. Since $Q(x)Q(x{+}d)>0$, the sign equality follows. Unless $Q(x)Q(x{+}d)$ is $x$-independent for each $d$, the gain cannot be written as an affine function of a finite-dimensional lift, so $F\notin\mathcal{DA}^{r}$ in general.
\end{proof}

This subsumes linear–fractional \cite{charnes1962programming}, polynomial–fractional \cite{dinkelbach1967nonlinear}, and finite-product energies \cite{ecker1980geometric, koller2009probabilistic}, although full DA holds only in the special cases noted above.

\subsection{Closure under composition}
\label{subsec:compositional}

Let $h_j\in\mathcal{DA}^{r_j}$ with feature maps $\psi_j$ so that for every EFD $d$, $\Delta_d h_j(x)=\left\langle W_{j,d},\psi_j(x)\right\rangle$.

\begin{proposition}[Affine and polynomial outer maps preserve DA]
\label{prop:dl-composition}
(a) If $f(x)=g\left(h_1(x),\dots,h_k(x)\right)$ with $g(z)=\alpha_0+\sum_{j=1}^k \alpha_j z_j$ affine, then with
\[
\Psi_{\oplus}(x):=\bigoplus_{j=1}^k \psi_j(x)\in\mathbb{R}^{\sum_j r_j}
\]
we have $\Delta_d f(x)=\left\langle W^{(\mathrm{aff})}_d,\ \Psi_{\oplus}(x)\right\rangle$ and $f\in\mathcal{DA}^{\sum_j r_j}$.\\
(b) If $f(x)=g\left(h_1(x),\dots,h_k(x)\right)$ with $g$ a polynomial of total degree $D\ge 1$, let $H(x):=(h_1(x),\dots,h_k(x))$ and let $\Phi_{\le D-1}(x)$ be the vector of all monomials in $H(x)$ of total degree $\le D-1$ (dimension $M=\binom{k+D-1}{D-1}$). Then
\[
\Delta_d f(x)=\left\langle W^{(\mathrm{poly})}_d,\ \Psi_{\oplus}(x)\otimes \Phi_{\le D-1}(x)\right\rangle,
\]
so $f\in\mathcal{DA}^{R}$ with $R\le \left(\sum_{j=1}^k r_j\right)\binom{k+D-1}{D-1}$.
\end{proposition}

\begin{proof}
(a) $\Delta_d f=\sum_j \alpha_j\Delta_d h_j$, each term linear in $\psi_j$; stack them. (b) Expand $g$ into monomials $\prod_{j} h_j^{\alpha_j}$ and apply the discrete product rule; every term contains exactly one factor $\Delta_d h_{j^\star}$ (linear in $\psi_{j^\star}$) times a product of at most $D-1$ remaining $h$’s, which is a coordinate of $\Phi_{\le D-1}(x)$.
\end{proof}

This subsumes hierarchical low-rank energies and shallow polynomial networks over DA primitives, including Boltzmann machines \cite{hinton1986learning} and energy-based models \cite{smolensky1986information, lecun2006tutorial}.

\subsection{Additively separable smooth functions}
\label{subsec:add-sep}

Consider $f(x)=\sum_{i=1}^n g_i\left(a_i^\top x\right)$ with $a_i\in\mathbb{R}^n$. Assume each $g_i$ admits a finite basis expansion of size $k_i$:
\[
\text{\emph{(poly)}}\quad g_i(z)=\sum_{\ell=0}^{k_i}\alpha_{i\ell} z^\ell
\qquad\text{or}\qquad
\text{\emph{(exp)}}\quad g_i(z)=\sum_{\ell=1}^{k_i}\alpha_{i\ell} e^{\beta_{i\ell} z}.
\]

\begin{proposition}[Finite-basis separable sums are DA]
\label{prop:add-sep-dl}
Let $\mathcal D=\{d^{(1)},\dots,d^{(m)}\}$ be the EFDs. Define $r:=\sum_{i=1}^n k_i$ and the feature map
\[
\psi(x)=
\begin{cases}
\left( \left(a_i^\top x\right)^\ell \right)_{i,\ \ell=0,\dots,k_i} & \text{\emph{(poly)}}\\[2pt]
\left( e^{\beta_{i\ell} a_i^\top x} \right)_{i,\ \ell=1,\dots,k_i} & \text{\emph{(exp)}}
\end{cases}
\quad\in\mathbb{R}^{r+(\text{const})}.
\]
Then for every $d\in\mathcal D$ there exists $w_d\in\left(\mathbb{R}^{r+(\text{const})}\right)^*$ such that $\Delta_d f(x)=\left\langle w_d,\psi(x)\right\rangle$. Hence $f\in\mathcal{DA}^{r}$.
\end{proposition}

\begin{proof}
\emph{(poly)} Let $s_i(x):=a_i^\top x$. Then
$
s_i(x{+}d)^\ell-s_i(x)^\ell=\sum_{t=0}^{\ell-1}\binom{\ell}{t}s_i(x)^t \left(a_i^\top d\right)^{\ell-t}
$,
linear in $\{s_i(x)^t\}_{t\le \ell-1}$. Summing over $(i,\ell)$ yields the claim.  
\emph{(exp)} $e^{\beta_{i\ell}\left(s_i(x)+a_i^\top d\right)}-e^{\beta_{i\ell}s_i(x)}=\left(e^{\beta_{i\ell}a_i^\top d}-1\right)e^{\beta_{i\ell}s_i(x)}$, linear in the exponential feature; summing over $(i,\ell)$ yields the claim.
\end{proof}

This subsumes finite polynomial truncations and Taylor series approximations \cite{rudin1987real}, mixtures of exponentials \cite{mclachlan2000finite}, finite-basis generalized linear model energies \cite{mccullagh2019generalized}.

\subsection{Bilinear and multilinear forms}
\label{subsec:bilinear}

\begin{proposition}[Low-rank bilinear and multilinear forms are DA]
\label{prop:bilinear-multilinear-da}
\textbf{(Bilinear)} Let $x\in\{0,1\}^n$, $y\in\{0,1\}^m$, and $M=\sum_{\ell=1}^r a_\ell b_\ell^\top$ with $a_\ell\in\mathbb{R}^n$, $b_\ell\in\mathbb{R}^m$. Consider
\[
f(x,y)=x^\top M y=\sum_{\ell=1}^r (a_\ell^\top x)(b_\ell^\top y).
\]
Define the feature map
\[
\psi(x,y)=\left((a_\ell^\top x)_{\ell=1}^r,\ (b_\ell^\top y)_{\ell=1}^r\right)\in\mathbb{R}^{2r}.
\]
For any EFD $d=(d_x,d_y)$, set
\[
v_d:=\left(\left(a_\ell^\top d_x\right)_{\ell=1}^r,\ \left(b_\ell^\top d_y\right)_{\ell=1}^r\right)\in(\mathbb{R}^{2r})^*,
\qquad
\beta_d:=\sum_{\ell=1}^r \left(a_\ell^\top d_x\right)\left(b_\ell^\top d_y\right).
\]
Then
\[
\Delta_d f(x,y)=\langle v_d,\psi(x,y)\rangle+\beta_d,
\]
hence $f\in\mathcal{DA}^{2r}$. If EFDs affect only one block at a time (flip in $x$ or in $y$), then $\beta_d=0$.

\textbf{(Multilinear)} Let $t\ge 2$, variables $x^{(s)}\in\{0,1\}^{n_s}$, and a CP-rank-$r$ decomposition
\[
f\left(x^{(1)},\dots,x^{(t)}\right)=\sum_{\ell=1}^r \prod_{s=1}^t \left\langle a^{(s)}_\ell, x^{(s)}\right\rangle,\qquad a^{(s)}_\ell\in\mathbb{R}^{n_s}.
\]
For each nonempty proper $T\subsetneq [t]$, let
\[
\phi_{\ell,T}\left(x^{(1)},\dots,x^{(t)}\right):=\prod_{s\in T}\left\langle a^{(s)}_\ell, x^{(s)}\right\rangle,
\]
and let $\psi$ collect all $\phi_{\ell,T}$ over $\ell\in[r]$ and $T\subsetneq [t],T\neq\emptyset$. Then for each EFD $d=(d^{(1)},\dots,d^{(t)})$ there exist $v_d,\beta_d$ with
\[
\Delta_d f=\left\langle v_d,\psi\right\rangle+\beta_d.
\]
Since there are $2^t-2$ nonempty proper subsets $T$ and $r$ CP terms, $f\in\mathcal{DA}^{r(2^t-2)}$. If EFDs affect at most one block $x^{(s)}$, it suffices to keep only $T=[t]\setminus\{s\}$, giving $f\in\mathcal{DA}^{rt}$.
\end{proposition}

\begin{proof}
\textbf{Bilinear:} With $f=\sum_{\ell=1}^r (a_\ell^\top x)(b_\ell^\top y)$ and $d=(d_x,d_y)$,
\[
\Delta_d f
=\sum_{\ell=1}^r\left[(a_\ell^\top d_x)(b_\ell^\top y)+(a_\ell^\top x)(b_\ell^\top d_y)+(a_\ell^\top d_x)(b_\ell^\top d_y)\right]
=\left\langle v_d,\psi(x,y)\right\rangle+\beta_d .
\]
\textbf{Multilinear:} Write $u^{(s)}_{\ell}(x):=\left\langle a^{(s)}_\ell,x^{(s)}\right\rangle$ and $\delta^{(s)}_{\ell}:=\left\langle a^{(s)}_\ell,d^{(s)}\right\rangle$. Then
\[
\Delta_d \left(\prod_{s=1}^t u^{(s)}_{\ell}\right) =\prod_{s=1}^t\left(u^{(s)}_{\ell}+\delta^{(s)}_{\ell}\right)-\prod_{s=1}^t u^{(s)}_{\ell} =\sum_{\emptyset\neq S\subseteq[t]}\ \left(\prod_{s\in S}\delta^{(s)}_{\ell}\right)\left(\prod_{s\notin S}u^{(s)}_{\ell}\right),
\]
an affine combination of partial products $\phi_{\ell,T}$ with $T=[t]\setminus S$ (the $S=[t]$ term yields $\beta_d$). Summing over $\ell$ yields the DA decomposition and the stated dimension bounds.
\end{proof}

This subsumes low-rank assignment \cite{lawler1963quadratic} and matrix factorization energies \cite{lee1999learning}, and CP-rank-$r$ multilinear objectives \cite{hitchcock1927expression}.

\section{Applications}
\label{sec:applications}

\subsection{Trimmed and robust statistics}
\label{sec:trimmed-robust}

See \cite{rousseeuw2011robust} for an introduction to robust statistics. 

\begin{remark}[Differentially linear relaxation]
\label{rem:dl-relaxation}
We write $f \in \mathcal{DL}^r$ (differentially linear) if discrete gains admit $\Delta_d f(x) = \langle w_d, \psi(x)\rangle / \Pi_d(x)$ where $\Pi_d(x) > 0$. Since signs are preserved, $\mathcal{DL}$ suffices for chamber enumeration. Note $\mathcal{DA}^r \subset \mathcal{DL}^r$.
\end{remark}

Let $X\in\mathbb{R}^{n\times d}$, $y\in\mathbb{R}^n$, and $x\in\{0,1\}^n$ select a subset via $W(x)=\mathrm{Diag}(x)$. Define the linear aggregates
\[
k=\mathbf{1}^\top x,\quad T_{rs}(x)=\sum_i x_i X_{ir}X_{is},\quad
s_r(x)=\sum_i x_i X_{ir}y_i,\quad
S_{yy}(x)=\sum_i x_i y_i^2,
\]
and collect $A(x):=\{k,S_{yy},T_{rs},s_r\}$ with $K:=|A|=\tfrac{d(d+1)}{2}+d+2$.

\begin{proposition}[Rational-in-aggregates are $\mathcal{DL}$]
\label{prop:rational-aggregates-dl}
Let $F(x)=P(A(x))/Q(A(x))$ where $P,Q$ are polynomials of total degrees $D_P,D_Q$ in the entries of $A(x)$, and assume $Q(A(x))>0$ on the feasible region and $\mathrm{rank}T(x)=d$ when $T^{-1}$ appears. Set $D:=D_P+D_Q-1$ and define the Veronese feature map \cite{matousek2013lectures} $\psi:\{0,1\}^n\to\mathbb{R}^{R}$ stacking all monomials in $A(x)$ of total degree $\le D$, with $R=\binom{K+D}{D}$. Then for every elementary feasible direction $d$ there exist $w_d\in\left(\mathbb{R}^{R}\right)^*$ and a positive $\Pi_d(x)$ such that
\[
\Delta_d F(x)=\frac{\langle w_d,\psi(x)\rangle}{\Pi_d(x)}.
\]
Hence $F\in\mathcal{DL}^{R}$, and exact optimization over any feasible family with a linear-optimization oracle runs in $O\left( m^r \right)$ byby \Cref{thm:xp-unconstrained}.
\end{proposition}

\begin{remark}
$\mathcal{DL}^R$ is the class of functions whose discrete gains admit an affine comparator up to a positive scaling factor, so $\mathcal{DA}^R\subset\mathcal{DL}^R$. This suffices for sign-consistent chamber enumeration.
\end{remark}

\begin{corollary}[Trimmed covariance]
\label{cor:trimmed-cov}
For $F(x)=\mathrm{Cov}\left(X^\top x,y^\top x\right)$ under fixed subset size $k=\mathbf{1}^\top x$, $F$ is linear–fractional in $A(x)$ with $(D_P,D_Q)=(2,2)$, giving $D=3$ and $R=\binom{K+3}{3}$; thus $F\in\mathcal{DL}^R$ and is optimizable in $O(T(n,m)m^{\binom{K+3}{3}})$ time.
\end{corollary}

\begin{corollary}[Trimmed $R^2$]
\label{cor:trimmed-r2}
For one-dimensional $y$ and $d$-dimensional $X$, $F(x)=R^2(x)=\mathrm{Cov}^2/\left(\mathrm{Var}_x\mathrm{Var}_y\right)$ is a ratio with $(D_P,D_Q)=(4,2)$, hence $D=5$ and $R=\binom{K+5}{5}$. Therefore $R^2\in\mathcal{DL}^R$ and admits $O\left(m^{\binom{K+5}{5}}\right)$ optimization.
\end{corollary}

\begin{corollary}[Trimmed BLUE slopes]
\label{cor:trimmed-blue}
For $\hat\beta(x)=\left(X^\top W X\right)^{-1}X^\top W y$,
each coordinate $\hat\beta_j(x)=P_j(A(x))/Q_j(A(x))$ with $D_P\le d+1$, $D_Q\le d$, so $D\le 2d$ and $R=\binom{K+2d}{2d}$; thus $\hat\beta_j\in\mathcal{DL}^R$.
\end{corollary}

\begin{corollary}[Trimmed Gauss–Markov RSS]
\label{cor:trimmed-rss}
$\mathrm{RSS}(x)=S_{yy}(x)-s(x)^\top T(x)^{-1}s(x)$ equals $P(A(x))/Q(A(x))$ with $D_P\le d+1$, $D_Q\le d$, hence $D\le 2d$ and $R=\binom{K+2d}{2d}$; therefore $\mathrm{RSS}\in\mathcal{DL}^R$.
\end{corollary}

\subsection{Optimal experimental design}\label{subsec:oed}

DA also applies to D-, A-, and E-optimality in optimal experimental design (OED) \cite{kiefer1959optimum}. Let $a_i\in\mathbb{R}^d$, $x\in\{0,1\}^n$, and $M(x)=\sum_{i=1}^n x_ia_i a_i^\top\in\mathbb{R}^{d\times d}$. Write the quadratic Veronese features \cite{matousek2013lectures} $\phi(x)=\operatorname{vecs}(M(x))\in\mathbb{R}^{K}$ with $K=d(d+1)/2$. For an integer $D\ge 0$, let $\psi_D(x)$ collect all monomials in the coordinates of $\phi(x)$ of total degree $\le D$ (including the constant $1$).

\begin{proposition}[Differential linearity of canonical OED criteria]
The following objectives are EFD-wise differentially linear with respect to suitable lifts $\psi_D$: (i) D-optimality $f_D(x)=\det M(x)$ with $D=d-1$; (ii) A-optimality $f_A(x)=-\operatorname{tr}(M(x)^{-1})=-\frac{\operatorname{adj}(M(x)):I}{\det M(x)}$ with $D=2d-2$; (iii) for any fixed $\tau\in\mathbb{R}$, the E-optimality decision polynomial $g_\tau(x)=\det(M(x)-\tau I)$ with $D=d-1$. Consequently, there exist covectors $W_d$ (one per elementary feasible direction $d$) such that for all feasible $x$ and EFDs $d$,
\[
\Delta_d f_D(x)=\left\langle W^{(D)}_d,\psi_{d-1}(x)\right\rangle,\qquad
\Delta_d f_A(x)=\frac{\left\langle W^{(A)}_d,\psi_{2d-2}(x)\right\rangle}{\det M(x)\det M(x{+}d)},\qquad
\Delta_d g_\tau(x)=\left\langle W^{(\tau)}_d,\psi_{d-1}(x)\right\rangle.
\]
In particular, $f_D\in\mathcal{DL}^{R_D}$, $f_A\in\mathcal{DL}^{R_A}$, and $g_\tau\in\mathcal{DL}^{R_E}$ with
\[
R_D=\binom{K+d-1}{d-1},\qquad R_A=\binom{K+2d-2}{2d-2},\qquad R_E=\binom{K+d-1}{d-1}.
\]
\end{proposition}

\begin{proof}[Proof sketch]
Each entry of $M$ is linear in $x$, so $\det M$ is a degree-$d$ polynomial in $\phi(x)$; for any EFD $d$, the finite difference $\det\left(M(x + d)\right)-\det(M(x))$ is a polynomial in $\phi(x)$ of degree $\le d-1$, hence a linear form in $\psi_{d-1}(x)$. For A-optimality, $\operatorname{adj}(M)$ and $\det(M)$ are degree $d-1$ and $d$ polynomials in $\phi(x)$; clearing denominators in $\Delta_d f_A$ yields a numerator polynomial of degree $\le (d-1)+(d-1)=2d-2$, i.e., a linear form in $\psi_{2d-2}(x)$, divided by the positive factor $\det M(x)\det M(x{+}d)$. For E-optimality in decision form, $g_\tau(x)=\det(M(x)-\tau I)$ is a degree-$d$ polynomial in $\phi(x)$, so its finite difference is linear in $\psi_{d-1}(x)$.
\end{proof}

\subsection{Portfolio optimization}
\label{subsec:portfolio}

Classical mean–variance portfolio selection \cite{markowitz1952modern} with factor risk \cite{sharpe1963simplified,fama1992cross} admits a $\mathcal{DL}$ reduction under equal-weighted subset selection. Let assets $i\in[n]$ have expected returns $\mu_i$, idiosyncratic volatilities $\sigma_i>0$, and $p$-factor loadings $b_{i\ell}$ with diagonal factor covariance $\Lambda=\mathrm{Diag}(\lambda_1,\dots,\lambda_p)\succ0$; the covariance is $\Sigma=B\Lambda B^\top+\mathrm{Diag}(\sigma_1^2,\dots,\sigma_n^2)$. For a selection $x\in\{0,1\}^n$ of size $k=\mathbf 1^\top x$, the equal-weight portfolio has mean
\[
\mu(x)=\frac{1}{k}\sum_{i} \mu_i x_i \quad\text{and}\quad
\mathrm{Var}(x)=\frac{1}{k^2}\left(\sum_{\ell=1}^p \lambda_\ell F_\ell(x)^2\;+\;D_{\mathrm{id}}(x)\right),
\]
where the \emph{portfolio aggregates} are linear in $x$:
\[
U(x):=\sum_i \mu_i x_i,\qquad F_\ell(x):=\sum_i b_{i\ell}x_i\ (\ell\in[p]),\qquad
S_\sigma(x):=\sum_i \sigma_i x_i,\qquad D_{\mathrm{id}}(x):=\sum_i \sigma_i^2 x_i.
\]
Collect $A(x):=\{k,U,F_1,\dots,F_p,S_\sigma,D_{\mathrm{id}}\}$ with $K=p+4$ (all linear in $x$).

\begin{proposition}[Rational-in-portfolio-aggregates are $\mathcal{DL}$]
\label{prop:portfolio-dl}
Let $F(x)=P(A(x))/Q(A(x))$ with $P,Q$ polynomials of total degrees $D_P,D_Q$ in the entries of $A(x)$, and assume $Q(A(x))>0$ on the feasible region (true here since $k\ge1$ and variances are positive). Set $D:=D_P+D_Q-1$ and let $\psi:\{0,1\}^n\to\mathbb{R}^{R}$ be the Veronese lift stacking all monomials in $A(x)$ of total degree $\le D$ \cite{matousek2013lectures}, with $R=\binom{K+D}{D}$. Then for every EFD $d$ there exist $w_d\in\left(\mathbb{R}^{R}\right)^*$ and a positive $\Pi_d(x)$ such that
\[
\Delta_d F(x)=\frac{\langle w_d,\psi(x)\rangle}{\Pi_d(x)}.
\]
Hence $F\in\mathcal{DL}^{R}$ and exact optimization over any feasible family with a linear-optimization oracle runs in $O\left(m^{R}\right)$ by \cref{thm:xp-unconstrained}.
\end{proposition}

\begin{corollary}[Mean–variance tradeoff]
\label{cor:mv}
For $F_{\mathrm{MV}}(x)=\alpha\mu(x)-\beta\mathrm{Var}(x)$ with $\alpha,\beta\ge0$, clearing denominators gives $F_{\mathrm{MV}}(x)=\frac{\alpha kU - \beta\left(\sum_\ell \lambda_\ell F_\ell^2 + D_{\mathrm{id}}\right)}{k^2},$ so $(D_P,D_Q)=(2,2)$ and $D=3$. Thus $F_{\mathrm{MV}}\in\mathcal{DL}^{R}$ with $R=\binom{K+3}{3}$ and admits $O\left(m^{\binom{K+3}{3}}\right)$ optimization.
\end{corollary}

\begin{corollary}[Squared Sharpe ratio]
\label{cor:sharpe}
The squared Sharpe ratio of the equal-weight portfolio is $S^2(x)=\mu(x)^2/\mathrm{Var}(x)=\frac{U^2}{\sum_\ell \lambda_\ell F_\ell^2 + D_{\mathrm{id}}},$ a ratio with $(D_P,D_Q)=(2,2)$, hence $D=3$ and $R=\binom{K+3}{3}$. Therefore $S^2\in\mathcal{DL}^{R}$ with the same complexity bound.
\end{corollary}

\begin{corollary}[Diversification ratio (squared)]
\label{cor:diversification}
The diversification ratio \cite{choueifaty2008towards,maillard2008properties} for equal weights is $ \mathrm{DR}(x)=\frac{\sum_i w_i\sigma_i}{\sqrt{w^\top \Sigma w}} =\frac{S_\sigma/k}{\sqrt{\mathrm{Var}(x)}}.$ Then $\mathrm{DR}^2(x)=\frac{S_\sigma^2}{\sum_\ell \lambda_\ell F_\ell^2 + D_{\mathrm{id}}}$ has $(D_P,D_Q)=(2,2)$, so $D=3$, $R=\binom{K+3}{3}$, and $\mathrm{DR}^2\in\mathcal{DL}^R$.
\end{corollary}

\section{Relation to existing paradigms}
\label{sec:existing-paradigms}

Our rank parameter serves as a structural width measure: bounding it yields fixed-parameter or XP-time tractability, analogous to treewidth in Courcelle-type meta-theorems \cite{elberfeld2016variants} and to Grohe’s structural parameters for constraint satisfaction \cite{grohe2007complexity}. In contrast to low-rank continuous relaxations for nonlinear programs, our approach is purely combinatorial, operating directly on the discrete hyperplane arrangement and producing exact solutions.

Prior to our work, even for QUBO, only limited classes of nonlinear objectives were known to be tractable, e.g., where $Q$ is positive semidefinite \cite{boumal2016nonconvex}, a sum matrix, a product matrix (particularly in the quadratic assignment problem \cite{cela2015well}), a Toeplitz matrix \cite{burkard1998quadratic}, has non-negative off-diagonal elements \cite{li2010polynomially, picard1982selected, picard1975minimum}, etc. Other polynomial solutions were limited to cases where $Q$ represents a tractable graph structure like a planar graph \cite{liers2012partitioning, lipton1979separator}, graph of bounded bandwidth \cite{dell1997annotated}, graph of bounded treewidth \cite{bodlaender2000complexity, bodlaender2008combinatorial}, graph of bounded genus \cite{galluccio2001optimization}, perfect graph \cite{grotschel1984polynomial}, t-perfect graphs \cite{grotschel1995combinatorial}, interaction graphs, \cite{boros2007pseudo}, etc. For Ising models, some prior work had explored separable energies \cite{barahona1982np,boros2002pseudo}. See \cite{punnen2022quadratic} for a complete list. 

Related work by Del-Pia and Khajavirad \cite{delpia2017polyhedral,delpia2023pseudoboolean,delpia2025factorized} gives polynomial-size extended formulations for binary polynomials with few multiplicative factors solved via mixed-integer programming. Our framework is complementary and more general in key respects, namely in subsuming such factorized models via suitable $\psi$ without convexification and yields direct combinatorial enumeration in $n^{O(r)}$ time. Solving polynomial programs by extending methods from quadratic programs had already been considered \cite{anthony2017quadratic}, albeit without the same XP guarantees. Dinklebach's method \cite{dinkelbach1967nonlinear} and Charnes-Cooper transform \cite{charnes1962programming} drew a connection between fractional programs and established methods in polynomial and linear programming, with guarantees of tractability in the latter case.

\section{Conclusion and outlook}
\label{sec:conclusion}

We introduced \emph{differential affineness} (DA) as a unifying structural parameter for nonlinear pseudo-Boolean optimization and proved an XP metatheorem: whenever an objective $f$ has DA rank $r$, maximizing $f$ reduces to enumerating $O(m^{r})$ chambers of an explicit affine hyperplane arrangement (where $m$ is the number of elementary directions), together with a two-sided local-optimality test yielding $O(n)$-time decoding per chamber (\Cref{thm:xp-unconstrained}). The framework is constructive: we provided closed-form lifts $\psi$, covectors $\{v_d\}$, and offsets $\{\beta_d\}$ for broad objective classes—low-rank quadratics, low-Waring-rank polynomials, products, compositions, separable sums, and CP-rank multilinear forms—and instantiated applications in robust statistics, experimental design, and portfolio optimization. Conceptually, DA rank plays a Courcelle-style width role for nonlinear objectives, replacing logical width (treewidth) with an analytic rank governing discrete gains.

\subsection{Limitations}
\begin{enumerate}
    \item \emph{Ratios and DL structure.} Fractional objectives $P/Q$ generally require directional comparisons when $Q$ varies; full DA holds only when $Q$ is constant along direction pairs.
    
    \item \emph{Rank blowups.} Composition and polynomial lifts can increase $r$ combinatorially (Veronese growth), preserving XP status but affecting practical constants.
    
    \item \emph{Constrained multiplicity.} With compound moves (e.g., swaps $\{\pm(e_i - e_j)\}$ for ranking), chambers may admit multiple stable points; uniqueness requires two-sided sign opposition at every direction.
    
    \item \emph{Model extraction.} Obtaining Waring or CP decompositions may be costly or approximate; our guarantees then apply to the extracted surrogate.
\end{enumerate}

\subsection{Open problems}
\begin{enumerate}
\item \emph{Constrained domains.} Extend to budget, knapsack, matroid, and exchange polytopes by characterizing $|\mathcal{D}|$ and exploiting arrangement sparsity.

\item \emph{Tighter rank bounds.} Close gaps between worst-case Veronese dimensions and problem-specific structure (symmetries, separability). Characterize minimal DA rank and its identifiability.

\item \emph{Tope-graph algorithms.} Replace full enumeration with output-sensitive walks exploiting oriented-matroid pivots, Gray-code sign updates, or incremental feature evaluation.

\item \emph{Learning $\psi$.} Given oracle access to $\Delta_d f(x)$, recover a minimal lift via structured low-rank regression with sample-complexity guarantees.

\item \emph{Smoothed analysis.} Prove that under random perturbations of $\{v_d, \beta_d\}$, opposite-sign conditions hold with high probability and the number of relevant chambers is polynomial.

\item \emph{Lower bounds.} Establish W[1]/ETH-style barriers when $r$ grows, or hardness when $\psi$ is unknown or only comparison oracles are available.
\end{enumerate}

DA rank isolates the combinatorial essence of nonlinear objectives: when discrete gains lie in a low-dimensional affine model, exact optimization becomes chamber enumeration in that dimension. This analytic width measure complements classical structural parameters (treewidth, rank-width) and provides a tractable, explicit route through problems previously requiring heavy relaxations. The framework's practical impact will grow with tighter rank bounds, extensions to constrained domains, and efficient tope-walk solvers.

\bibliographystyle{ACM-Reference-Format}
\bibliography{refs}

\end{document}